\documentclass[12pt, english]{article}

\usepackage{babel}
\usepackage[latin1]{inputenc}
\usepackage{amsmath}
\usepackage{amssymb}
\usepackage{amsthm}
\usepackage{graphicx}

\theoremstyle{plain}
\newtheorem{Thm}{Theorem}%[section]
\newtheorem{Lemma}[Thm]{Lemma}

\usepackage{pgf}
\usepackage{tikz}
\usetikzlibrary{arrows}
\usetikzlibrary{decorations.pathmorphing}
\usetikzlibrary{backgrounds}
\usetikzlibrary{fit}

\title{From heaps of matches to the limits of computability}
\author{Urban Larsson and Johan W\"astlund\\ Mathematical Sciences,\\ Chalmers University of Technology and University of Gothenburg,\\ G\"oteborg, Sweden\\
urban.larsson@chalmers.se, wastlund@chalmers.se}
\date{\today}

\begin{document}

\maketitle

\begin{abstract}
We study so-called invariant games played with a fixed number $d$ of heaps of matches. A game is described by a finite list $\mathcal{M}$ of integer vectors of length $d$ specifying the legal moves. A move consists in changing the current game-state by adding one of the vectors in $\mathcal{M}$, provided all elements of the resulting vector are nonnegative. 
For instance, in a two-heap game, the vector $(1,-2)$ would mean adding one match to the first heap and removing two matches from the second heap. If $(1,-2) \in \mathcal{M}$, such a move would be permitted provided there are at least two matches in the second heap.   
Two players take turns, and a player unable to make a move loses. 
We show that these games embrace computational universality, and that therefore a number of basic questions about them are algorithmically undecidable. In particular, we prove that there is no algorithm that takes two games $\mathcal{M}$ and $\mathcal{M}'$ (with the same number of heaps) as input, and determines whether or not they are equivalent in the sense that every starting-position which is a first player win in one of the games is a first player win in the other.\end{abstract}

\section{Introduction}
\subsection{A children's game}
A type of children's game for two players consists in placing a heap of matches on a table and taking turns removing them according to some simple rule, the winner being the person to make the last move. For instance, the rule can be that one is permitted to remove one, two or three matches. Playing a few games will lead to the insight that certain positions are advantageous in the sense that moving to them will secure the win. In this example, moving to a heap of four matches will secure the win in the next move. Similarly, a player moving to a heap of eight will be able to move to four in the next move, and in general, the \emph{P-positions} are precisely the multiples of four.

If the set of numbers that one is allowed to remove from the heap is finite, then the set of P-positions will ultimately become periodic. Therefore any particular game of this type can be completely understood. Finding the period may require a tedious computation, but the game cannot in principle embrace any mysteries, see \cite{BCG04}.

\subsection{Games of more than one heap}
We study similar games with several heaps of matches. Each game has a fixed number $d$ of heaps, and a position can be regarded as a $d$-dimensional vector $a=(a_1,\dots, a_d)$ of non-negative integers.
The rules of the game are encoded by a finite set $\mathcal{M}$ of integer vectors that specify the permitted moves. If $(m_1,\dots, m_d)\in \mathcal{M}$, then from position $(a_1,\dots,a_d)$, a player can move to position $(a_1+m_1,\dots,a_d+m_d)$, provided all of the numbers $a_1+m_1,\dots, a_d+m_d$ are nonnegative. These games are called \emph{invariant} games 
\cite{DR10, G66, L12, LHF11}, since (apart from the nonnegativity condition) the move options given by $\mathcal{M}$ are independent of $a$. This is somehow implicit in the idea of heaps of matches --- we should not have to count the remaining matches in order to play by the rules. 

Since moves can involve adding matches to heaps, the game does not necessarily have to terminate. We restrict our attention to games where termination is not an issue by stipulating that for each 
$(m_1,\dots, m_d) \in \mathcal{M}$, $m_1+\dots+m_d < 0$, so that in each move the total number of matches decreases.

Each game $\mathcal{M}$ has a set $P(\mathcal{M})$ of P-positions, and in informal terms our question is whether from the list $\mathcal{M}$ of permitted moves it is possible to completely understand $P(\mathcal{M})$. 

\begin{figure} [h]
\begin{center}
\begin{tikzpicture} [scale = 0.2]

\foreach \x/\y in {0/0,0/1,0/2,0/3,0/4,0/5,0/6,0/7,0/8,0/9,0/10,0/11,0/12,0/13,0/14,0/15,0/16,0/17,0/18,0/19,0/20,0/21,0/22,0/23,0/24,1/0,1/1,1/2,3/2,3/3,3/4,3/5,3/6,3/7,3/8,3/9,3/10,3/11,3/12,3/13,3/14,3/15,3/16,3/17,3/18,3/19,3/20,3/21,3/22,3/23,3/24,4/0,4/1,4/2,4/3,4/4,5/0,6/4,6/5,6/6,6/7,6/8,6/9,6/10,6/11,6/12,6/13,6/14,6/15,6/16,6/17,6/18,6/19,6/20,6/21,6/22,6/23,6/24,7/0,7/1,7/2,7/3,7/4,7/5,7/6,8/0,8/1,8/2,9/6,9/7,9/8,9/9,9/10,9/11,9/12,9/13,9/14,9/15,9/16,9/17,9/18,9/19,9/20,9/21,9/22,9/23,9/24,10/2,10/3,10/4,10/5,10/6,10/7,10/8,11/0,11/1,11/2,11/3,11/4,12/0,12/8,12/9,12/10,12/11,12/12,12/13,12/14,12/15,12/16,12/17,12/18,12/19,12/20,12/21,12/22,12/23,12/24,13/4,13/5,13/6,13/7,13/8,13/9,13/10,14/0,14/1,14/2,14/3,14/4,14/5,14/6,15/0,15/1,15/2,15/10,15/11,15/12,15/13,15/14,15/15,15/16,15/17,15/18,15/19,15/20,15/21,15/22,15/23,15/24,16/6,16/7,16/8,16/9,16/10,16/11,16/12,17/2,17/3,17/4,17/5,17/6,17/7,17/8,18/0,18/1,18/2,18/3,18/4,18/12,18/13,18/14,18/15,18/16,18/17,18/18,18/19,18/20,18/21,18/22,18/23,18/24,19/0,19/8,19/9,19/10,19/11,19/12,19/13,19/14,20/4,20/5,20/6,20/7,20/8,20/9,20/10,21/0,21/1,21/2,21/3,21/4,21/5,21/6,21/14,21/15,21/16,21/17,21/18,21/19,21/20,21/21,21/22,21/23,21/24,22/0,22/1,22/2,22/10,22/11,22/12,22/13,22/14,22/15,22/16,23/6,23/7,23/8,23/9,23/10,23/11,23/12,24/2,24/3,24/4,24/5,24/6,24/7,24/8,24/16,24/17,24/18,24/19,24/20,24/21,24/22,24/23,24/24}
 \filldraw[color = red] (\x, \y) rectangle (\x+1, \y+1);

\draw[step=1cm,gray,very thin] (0, 0) grid (25,25); 

\draw (0.5 ,-1.2) node {$0$};
\draw (10.5, -1.2) node {$10$};
\draw (20.5, -1.2) node {$20$};

\draw (-1, 0.5) node {$0$};
\draw (-1.5, 10.5) node {$10$};
\draw (-1.5, 20.5) node {$20$};

\draw (12, -4) node {Heap 1};
\draw (-7, 13) node {Heap 2};

\end{tikzpicture}
\end{center}

\caption{The P-positions of the game $\mathcal{M} = \{(-1,-3), (-2, 1)\}$.}
\label{F:firstExample}
\end{figure}
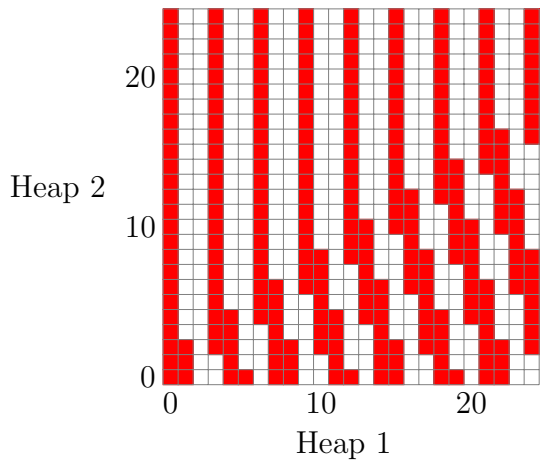

A first observation is that for any given position $a = (a_1,\dots,a_d)$ we can determine recursively whether or not $a$ is a P-position by first computing the status of all positions with fewer matches. The position $a$ is in $P(\mathcal{M})$ iff there is no move from $a$ to a position in $P(\mathcal{M})$. 
Consider for instance the game of two heaps given by $\mathcal{M} = \{(-1,-3), (-2, 1)\}$. The P-positions with fewer than 25 matches in each heap are shown in red in Figure~\ref{F:firstExample}.

There are two sectors, each with its own periodic pattern of P-positions, and clearly it would be possible to describe these patterns explicitly and prove by induction that they will persist. Therefore it is fair to say that we have complete understanding of the set of P-positions in this game.

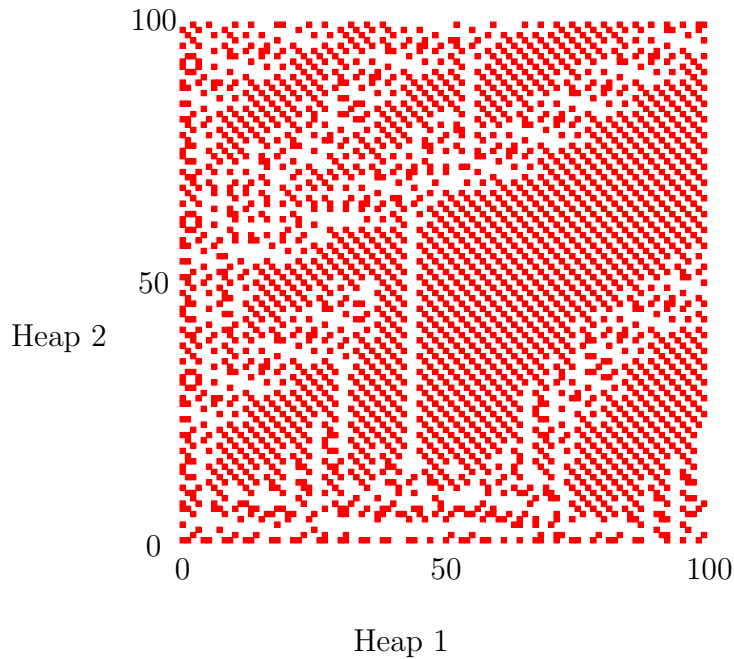
\begin{figure} [ht!]
\begin{center}
\begin{tikzpicture} [scale = 0.07]
\foreach \x/\y in {
0/1,0/4,0/7,0/8,0/11,0/14,0/15,0/18,0/21,0/24,0/27,0/28,0/31,0/32,0/35,0/38,0/41,0/44,0/45,0/48,0/51,0/54,0/57,0/58,0/61,0/62,0/65,0/68,0/71,0/74,0/75,0/78,0/81,0/84,0/87,0/88,0/91,0/92,0/95,0/98,1/1,1/7,1/10,1/13,1/16,1/17,1/20,1/21,1/24,1/27,1/30,1/33,1/37,1/40,1/43,1/46,1/47,1/50,1/51,1/54,1/57,1/60,1/63,1/67,1/70,1/73,1/76,1/77,1/80,1/81,1/84,1/87,1/90,1/93,1/97,2/2,2/6,2/9,2/10,2/13,2/16,2/19,2/22,2/30,2/33,2/36,2/39,2/40,2/43,2/46,2/49,2/52,2/60,2/63,2/66,2/69,2/70,2/73,2/76,2/79,2/82,2/90,2/93,2/96,2/99,3/3,3/8,3/11,3/28,3/31,3/32,3/35,3/38,3/41,3/58,3/61,3/62,3/65,3/68,3/71,3/88,3/91,3/92,3/95,3/98,4/1,4/20,4/23,4/26,4/29,4/50,4/53,4/56,4/59,4/80,4/83,4/86,4/89,5/1,5/12,5/15,5/18,5/21,5/39,5/42,5/45,5/48,5/51,5/69,5/72,5/75,5/78,5/81,5/99,6/7,6/8,6/11,6/14,6/27,6/28,6/31,6/34,6/37,6/40,6/57,6/61,6/64,6/67,6/68,6/71,6/74,6/87,6/88,6/91,6/94,6/97,7/3,7/7,7/10,7/13,7/16,7/19,7/22,7/25,7/28,7/46,7/49,7/52,7/55,7/58,7/73,7/76,7/79,7/82,7/85,7/88,8/1,8/6,8/9,8/12,8/15,8/18,8/21,8/35,8/38,8/41,8/44,8/47,8/48,8/51,8/52,8/55,8/75,8/78,8/81,8/95,8/98,9/1,9/14,9/17,9/20,9/23,9/27,9/30,9/33,9/34,9/37,9/40,9/41,9/44,9/47,9/50,9/53,9/60,9/63,9/64,9/67,9/68,9/71,9/74,9/77,9/80,9/83,9/87,9/90,9/93,9/94,9/97,10/2,10/7,10/13,10/16,10/19,10/22,10/25,10/28,10/33,10/36,10/39,10/42,10/58,10/59,10/62,10/63,10/66,10/69,10/73,10/76,10/79,10/82,10/85,10/88,10/93,10/96,10/99,11/5,11/8,11/12,11/15,11/18,11/21,11/51,11/54,11/75,11/78,11/81,12/1,12/8,12/11,12/14,12/17,12/20,12/23,12/26,12/29,12/34,12/37,12/40,12/43,12/46,12/49,12/52,12/64,12/67,12/68,12/71,12/74,12/77,12/80,12/83,12/86,12/89,12/94,12/97,13/1,13/6,13/7,13/10,13/13,13/16,13/19,13/22,13/25,13/28,13/32,13/35,13/39,13/42,13/45,13/58,13/59,13/62,13/65,13/69,13/79,13/82,13/85,13/88,13/92,13/95,13/99,14/2,14/7,14/12,14/15,14/18,14/21,14/24,14/27,14/41,14/44,14/47,14/50,14/53,14/58,14/78,14/81,14/84,14/87,15/5,15/17,15/20,15/23,15/26,15/33,15/34,15/37,15/40,15/43,15/46,15/49,15/52,15/63,15/67,15/70,15/73,15/74,15/77,15/80,15/83,15/86,15/93,15/94,15/97,16/6,16/13,16/16,16/19,16/22,16/25,16/28,16/31,16/34,16/35,16/45,16/48,16/51,16/54,16/59,16/62,16/65,16/66,16/69,16/72,16/73,16/76,16/79,16/82,16/85,16/88,16/91,16/94,16/95,17/1,17/6,17/11,17/12,17/15,17/18,17/21,17/24,17/27,17/41,17/44,17/47,17/50,17/53,17/57,17/81,17/84,17/87,18/1,18/4,18/7,18/11,18/14,18/17,18/20,18/23,18/26,18/29,18/33,18/36,18/39,18/40,18/43,18/46,18/49,18/52,18/74,18/77,18/80,18/83,18/86,18/89,18/93,18/96,18/99,19/7,19/10,19/13,19/16,19/19,19/22,19/25,19/28,19/31,19/34,19/39,19/42,19/45,19/48,19/51,19/58,19/59,19/62,19/65,19/68,19/72,19/75,19/82,19/85,19/88,19/91,19/94,19/99,20/5,20/6,20/21,20/24,20/27,20/47,20/50,20/53,20/56,20/60,20/67,20/78,20/81,20/84,20/87,21/1,21/6,21/17,21/20,21/23,21/26,21/29,21/32,21/35,21/40,21/43,21/46,21/49,21/52,21/63,21/66,21/70,21/73,21/76,21/77,21/80,21/83,21/86,21/89,21/92,21/95,22/1,22/9,22/12,22/15,22/16,22/19,22/22,22/25,22/28,22/31,22/34,22/38,22/41,22/45,22/48,22/51,22/54,22/58,22/61,22/62,22/65,22/68,22/71,22/76,22/79,22/82,22/85,22/88,22/91,22/94,22/98,23/2,23/8,23/11,23/12,23/15,23/18,23/21,23/24,23/27,23/30,23/33,23/47,23/50,23/53,23/56,23/59,23/87,23/90,23/93,24/5,24/10,24/11,24/14,24/17,24/20,24/23,24/26,24/29,24/32,24/39,24/40,24/43,24/46,24/49,24/52,24/66,24/69,24/72,24/80,24/83,24/86,24/89,24/92,24/99,25/3,25/6,25/25,25/28,25/31,25/34,25/37,25/40,25/41,25/51,25/54,25/57,25/61,25/64,25/65,25/68,25/71,25/75,25/78,25/81,25/82,25/85,25/88,25/91,25/94,25/97,26/1,26/27,26/30,26/33,26/47,26/50,26/53,26/56,26/70,26/84,26/87,26/90,26/93,27/1,27/2,27/7,27/10,27/13,27/16,27/19,27/20,27/23,27/26,27/29,27/32,27/35,27/39,27/42,27/45,27/46,27/49,27/52,27/55,27/66,27/69,27/77,27/80,27/83,27/86,27/89,27/92,27/95,27/99,28/12,28/15,28/18,28/21,28/22,28/25,28/28,28/31,28/34,28/37,28/40,28/45,28/48,28/51,28/54,28/57,28/60,28/64,28/65,28/68,28/71,28/74,28/77,28/78,28/94,28/97,29/8,29/11,29/14,29/15,29/18,29/21,29/24,29/27,29/30,29/33,29/53,29/56,29/59,29/70,29/73,29/87,29/90,29/93,30/1,30/6,30/7,30/10,30/13,30/16,30/38,30/41,30/46,30/49,30/52,30/55,30/58,30/72,30/76,30/79,30/82,30/85,30/88,30/89,30/92,30/95,30/98,31/1,31/2,31/6,31/7,31/36,31/39,31/44,31/47,31/51,31/54,31/57,31/60,31/64,31/65,31/68,31/71,31/74,31/77,31/82,31/85,31/88,31/91,31/94,31/97,32/8,32/14,32/17,32/20,32/23,32/26,32/29,32/32,32/50,32/53,32/56,32/59,32/66,32/70,32/83,32/96,32/99,33/9,33/13,33/16,33/19,33/22,33/25,33/28,33/31,33/34,33/37,33/40,33/43,33/46,33/49,33/52,33/55,33/58,33/72,33/75,33/78,33/81,33/84,33/89,33/92,33/95,33/98,34/1,34/6,34/12,34/15,34/18,34/21,34/24,34/27,34/30,34/33,34/36,34/39,34/43,34/46,34/51,34/54,34/57,34/60,34/64,34/65,34/68,34/71,34/74,34/77,34/81,34/84,34/87,34/90,34/94,34/97,35/1,35/7,35/11,35/14,35/17,35/20,35/23,35/26,35/29,35/32,35/35,35/38,35/41,35/44,35/53,35/56,35/59,35/62,35/65,35/66,35/76,35/79,35/82,35/96,35/99,36/13,36/16,36/19,36/22,36/25,36/28,36/31,36/34,36/37,36/52,36/55,36/58,36/72,36/75,36/88,36/89,36/92,36/95,36/98,37/5,37/6,37/9,37/21,37/24,37/27,37/30,37/33,37/36,37/39,37/42,37/45,37/48,37/51,37/54,37/57,37/60,37/64,37/67,37/70,37/71,37/74,37/77,37/80,37/83,37/86,37/89,37/90,38/1,38/6,38/7,38/14,38/17,38/20,38/23,38/26,38/29,38/32,38/35,38/38,38/41,38/44,38/47,38/50,38/53,38/56,38/59,38/62,38/65,38/70,38/73,38/76,38/79,38/82,38/96,38/99,39/1,39/7,39/12,39/15,39/16,39/19,39/22,39/25,39/28,39/31,39/34,39/37,39/40,39/43,39/46,39/49,39/52,39/55,39/58,39/78,39/81,39/84,39/88,39/91,39/94,39/95,39/98,40/8,40/12,40/15,40/18,40/21,40/24,40/27,40/30,40/33,40/36,40/39,40/42,40/45,40/48,40/51,40/54,40/57,40/60,40/63,40/66,40/71,40/74,40/77,40/80,40/83,40/86,40/89,40/94,40/97,41/6,41/10,41/11,41/14,41/17,41/20,41/23,41/26,41/29,41/32,41/35,41/38,41/41,41/44,41/47,41/50,41/53,41/56,41/59,41/62,41/65,41/66,41/69,41/72,41/76,41/79,41/82,42/6,42/19,42/22,42/25,42/28,42/31,42/34,42/37,42/40,42/43,42/46,42/49,42/52,42/55,42/58,42/61,42/64,42/78,42/81,42/84,42/87,42/90,42/95,42/98,43/1,43/5,43/12,43/70,43/71,43/74,43/77,43/80,43/83,43/86,43/89,43/93,43/96,44/1,44/7,44/10,44/13,44/65,44/68,44/71,44/72,44/82,44/85,44/88,45/6,45/7,45/18,45/21,45/24,45/27,45/30,45/33,45/36,45/39,45/42,45/45,45/48,45/51,45/54,45/57,45/60,45/63,45/66,45/81,45/84,45/87,45/94,45/95,45/98,46/5,46/14,46/17,46/20,46/23,46/26,46/29,46/32,46/35,46/38,46/41,46/44,46/47,46/50,46/53,46/56,46/59,46/73,46/76,46/77,46/80,46/83,46/86,46/89,46/92,46/95,46/96,47/1,47/12,47/13,47/16,47/19,47/22,47/25,47/28,47/31,47/34,47/37,47/40,47/43,47/46,47/49,47/52,47/55,47/58,47/61,47/64,47/67,47/71,47/72,47/75,47/76,47/79,47/82,47/85,47/88,48/6,48/9,48/12,48/15,48/18,48/21,48/24,48/27,48/30,48/33,48/36,48/39,48/42,48/45,48/48,48/51,48/54,48/57,48/60,48/63,48/69,48/84,48/87,48/90,48/94,48/97,49/7,49/8,49/11,49/14,49/17,49/20,49/23,49/26,49/29,49/32,49/35,49/38,49/41,49/44,49/47,49/50,49/53,49/56,49/59,49/62,49/65,49/77,49/80,49/83,49/86,49/89,49/92,49/95,50/1,50/5,50/13,50/16,50/19,50/22,50/25,50/28,50/31,50/34,50/37,50/40,50/43,50/46,50/49,50/52,50/55,50/58,50/61,50/64,50/67,50/71,50/72,50/75,50/78,50/85,50/88,51/2,51/6,51/12,51/18,51/21,51/24,51/27,51/30,51/33,51/36,51/39,51/42,51/45,51/48,51/51,51/54,51/57,51/60,51/63,51/66,51/84,51/87,51/90,51/93,51/96,52/7,52/11,52/17,52/20,52/23,52/26,52/29,52/32,52/35,52/38,52/41,52/44,52/47,52/50,52/53,52/56,52/59,52/62,52/65,52/76,52/79,52/80,52/83,52/86,52/89,52/92,52/95,52/96,52/99,53/1,53/7,53/13,53/16,53/19,53/22,53/25,53/28,53/31,53/34,53/37,53/40,53/43,53/46,53/49,53/52,53/55,53/58,53/61,53/64,53/67,53/72,53/75,53/78,53/79,53/82,53/85,53/88,53/91,53/94,54/1,54/6,54/12,54/15,54/18,54/21,54/24,54/27,54/30,54/33,54/36,54/39,54/42,54/45,54/48,54/51,54/54,54/57,54/60,54/63,54/66,54/69,54/72,55/2,55/6,55/14,55/17,55/20,55/23,55/26,55/29,55/32,55/35,55/38,55/41,55/44,55/47,55/50,55/53,55/56,55/59,55/62,55/65,55/95,55/98,56/7,56/16,56/19,56/22,56/25,56/28,56/31,56/34,56/37,56/40,56/43,56/46,56/49,56/52,56/55,56/58,56/61,56/64,56/67,56/71,56/74,56/75,56/78,56/81,56/84,56/87,56/90,56/93,56/96,57/1,57/5,57/8,57/12,57/15,57/18,57/21,57/24,57/27,57/30,57/33,57/36,57/39,57/42,57/45,57/48,57/51,57/54,57/57,57/60,57/63,57/66,57/69,57/74,57/77,57/80,57/83,57/86,57/89,58/1,58/6,58/10,58/11,58/14,58/17,58/20,58/23,58/26,58/29,58/32,58/35,58/38,58/41,58/44,58/47,58/50,58/53,58/56,58/59,58/62,58/65,58/82,58/85,58/88,58/91,58/94,58/97,59/6,59/7,59/19,59/22,59/25,59/28,59/31,59/34,59/37,59/40,59/43,59/46,59/49,59/52,59/55,59/58,59/61,59/64,59/67,59/70,59/75,59/78,59/81,59/84,59/87,59/90,59/93,59/99,60/5,60/12,60/15,60/18,60/21,60/24,60/27,60/30,60/33,60/36,60/39,60/42,60/45,60/48,60/51,60/54,60/57,60/60,60/63,60/66,60/69,60/73,60/76,60/80,60/83,60/86,60/89,60/92,60/95,61/1,61/5,61/10,61/13,61/20,61/23,61/26,61/29,61/32,61/35,61/38,61/41,61/44,61/47,61/50,61/53,61/56,61/59,61/62,61/65,61/68,61/82,61/85,61/88,61/91,61/94,61/97,62/6,62/22,62/25,62/28,62/31,62/34,62/37,62/40,62/43,62/46,62/49,62/52,62/55,62/58,62/61,62/64,62/67,62/74,62/75,62/78,62/81,62/84,62/87,62/90,62/93,62/96,63/4,63/11,63/14,63/15,63/18,63/21,63/24,63/27,63/30,63/33,63/36,63/39,63/42,63/45,63/48,63/51,63/54,63/57,63/60,63/63,63/66,63/69,63/72,63/75,63/76,63/86,63/89,63/92,63/95,64/1,64/4,64/5,64/10,64/13,64/16,64/17,64/20,64/23,64/26,64/29,64/32,64/35,64/38,64/41,64/44,64/47,64/50,64/53,64/56,64/59,64/62,64/65,64/68,64/82,64/85,64/88,64/91,64/94,64/97,65/6,65/10,65/31,65/34,65/37,65/40,65/43,65/46,65/49,65/52,65/55,65/58,65/61,65/64,65/67,65/70,65/74,65/77,65/80,65/81,65/84,65/87,65/90,65/93,65/96,65/99,66/8,66/11,66/33,66/36,66/39,66/42,66/45,66/48,66/51,66/54,66/57,66/60,66/63,66/66,66/69,66/72,66/75,66/80,66/83,66/86,66/89,66/92,66/95,67/1,67/4,67/5,67/9,67/16,67/19,67/22,67/25,67/26,67/29,67/32,67/35,67/38,67/41,67/44,67/47,67/50,67/53,67/56,67/59,67/62,67/65,67/68,67/88,67/91,67/94,67/97,68/1,68/2,68/15,68/18,68/21,68/24,68/27,68/28,68/31,68/34,68/37,68/40,68/43,68/46,68/49,68/52,68/55,68/58,68/61,68/64,68/67,68/70,68/73,68/76,68/81,68/84,68/87,68/90,68/93,68/96,68/99,69/10,69/11,69/14,69/17,69/20,69/21,69/24,69/27,69/30,69/33,69/36,69/39,69/42,69/45,69/48,69/51,69/54,69/57,69/60,69/63,69/66,69/69,69/72,69/75,69/79,69/82,69/86,69/89,69/92,69/95,70/4,70/9,70/10,70/13,70/16,70/19,70/22,70/38,70/41,70/44,70/47,70/50,70/53,70/56,70/59,70/62,70/65,70/68,70/71,70/74,70/88,70/91,70/94,70/97,71/1,71/2,71/5,71/6,71/9,71/31,71/34,71/37,71/40,71/43,71/46,71/49,71/52,71/55,71/58,71/61,71/64,71/67,71/70,71/73,71/80,71/81,71/84,71/87,71/90,71/93,71/96,71/99,72/2,72/20,72/23,72/26,72/29,72/32,72/33,72/36,72/39,72/42,72/45,72/48,72/51,72/54,72/57,72/60,72/63,72/66,72/69,72/72,72/75,72/78,72/81,72/82,72/92,72/95,72/98,73/12,73/15,73/18,73/21,73/35,73/38,73/41,73/44,73/47,73/50,73/53,73/56,73/59,73/62,73/65,73/68,73/71,73/74,73/88,73/91,73/94,73/97,74/1,74/5,74/8,74/11,74/14,74/28,74/31,74/34,74/37,74/40,74/43,74/46,74/49,74/52,74/55,74/58,74/61,74/64,74/67,74/70,74/73,74/76,74/80,74/83,74/86,74/87,74/90,74/93,74/96,74/99,75/1,75/7,75/10,75/13,75/16,75/19,75/22,75/25,75/28,75/29,75/45,75/48,75/51,75/54,75/57,75/60,75/63,75/66,75/69,75/72,75/75,75/78,75/81,75/86,75/89,75/92,75/95,75/98,76/6,76/9,76/12,76/15,76/18,76/21,76/24,76/38,76/41,76/44,76/47,76/50,76/53,76/56,76/59,76/62,76/65,76/68,76/71,76/74,76/94,76/97,77/8,77/11,77/14,77/17,77/20,77/23,77/27,77/30,77/33,77/36,77/39,77/40,77/43,77/46,77/49,77/52,77/55,77/58,77/61,77/64,77/67,77/70,77/73,77/76,77/79,77/82,77/87,77/90,77/93,77/96,77/99,78/1,78/7,78/10,78/13,78/16,78/19,78/22,78/25,78/28,78/33,78/36,78/39,78/42,78/45,78/48,78/51,78/54,78/57,78/60,78/63,78/66,78/69,78/72,78/75,78/78,78/81,78/85,78/88,78/92,78/95,78/98,79/2,79/6,79/9,79/12,79/15,79/18,79/21,79/34,79/47,79/50,79/53,79/56,79/59,79/62,79/65,79/68,79/71,79/74,79/77,79/80,79/94,79/97,80/5,80/8,80/11,80/14,80/17,80/20,80/23,80/26,80/29,80/32,80/35,80/40,80/43,80/46,80/49,80/52,80/55,80/58,80/61,80/64,80/67,80/70,80/73,80/76,80/79,80/86,80/87,80/90,80/93,80/96,80/99,81/1,81/7,81/10,81/13,81/16,81/19,81/22,81/25,81/28,81/32,81/35,81/38,81/41,81/45,81/48,81/51,81/54,81/57,81/60,81/63,81/66,81/69,81/72,81/75,81/78,81/81,81/84,81/87,81/88,81/98,82/1,82/6,82/9,82/12,82/15,82/18,82/21,82/24,82/27,82/30,82/33,82/47,82/50,82/53,82/56,82/59,82/62,82/65,82/68,82/71,82/74,82/77,82/80,82/94,82/97,83/2,83/11,83/14,83/17,83/20,83/23,83/26,83/39,83/40,83/43,83/46,83/49,83/52,83/55,83/58,83/61,83/64,83/67,83/70,83/73,83/76,83/79,83/82,83/86,83/89,83/92,83/93,83/96,83/99,84/7,84/10,84/13,84/16,84/19,84/22,84/25,84/28,84/31,84/34,84/37,84/40,84/41,84/51,84/54,84/57,84/60,84/63,84/66,84/69,84/72,84/75,84/78,84/81,84/84,84/87,84/92,84/95,84/98,85/5,85/6,85/9,85/12,85/15,85/18,85/21,85/24,85/27,85/30,85/33,85/47,85/50,85/53,85/56,85/59,85/62,85/65,85/68,85/71,85/74,85/77,85/80,86/1,86/5,86/8,86/11,86/14,86/17,86/20,86/23,86/26,86/29,86/32,86/35,86/39,86/42,86/45,86/46,86/49,86/52,86/55,86/58,86/61,86/64,86/67,86/70,86/73,86/76,86/79,86/82,86/85,86/88,86/93,86/96,86/99,87/1,87/4,87/7,87/10,87/13,87/16,87/19,87/22,87/25,87/28,87/31,87/34,87/37,87/40,87/45,87/48,87/51,87/54,87/57,87/60,87/63,87/66,87/69,87/72,87/75,87/78,87/81,87/84,87/87,87/91,87/94,87/98,88/15,88/18,88/21,88/24,88/27,88/30,88/33,88/53,88/56,88/59,88/62,88/65,88/68,88/71,88/74,88/77,88/80,88/83,88/86,89/11,89/14,89/17,89/20,89/23,89/26,89/29,89/32,89/35,89/38,89/41,89/46,89/49,89/52,89/55,89/58,89/61,89/64,89/67,89/70,89/73,89/76,89/79,89/82,89/85,89/92,89/93,89/96,89/99,90/3,90/6,90/9,90/10,90/13,90/16,90/19,90/22,90/25,90/28,90/31,90/34,90/37,90/40,90/44,90/47,90/51,90/54,90/57,90/60,90/63,90/66,90/69,90/72,90/75,90/78,90/81,90/84,90/87,90/90,90/93,90/94,91/2,91/5,91/6,91/9,91/12,91/15,91/18,91/21,91/24,91/27,91/30,91/33,91/36,91/39,91/53,91/56,91/59,91/62,91/65,91/68,91/71,91/74,91/77,91/80,91/83,91/86,92/1,92/2,92/5,92/8,92/11,92/14,92/17,92/20,92/23,92/26,92/29,92/32,92/35,92/38,92/45,92/46,92/49,92/52,92/55,92/58,92/61,92/64,92/67,92/70,92/73,92/76,92/79,92/82,92/85,92/88,92/92,92/95,92/98,92/99,93/19,93/22,93/25,93/28,93/31,93/34,93/37,93/40,93/43,93/46,93/47,93/57,93/60,93/63,93/66,93/69,93/72,93/75,93/78,93/81,93/84,93/87,93/90,93/93,93/98,94/18,94/21,94/24,94/27,94/30,94/33,94/36,94/39,94/53,94/56,94/59,94/62,94/65,94/68,94/71,94/74,94/77,94/80,94/83,94/86,95/1,95/4,95/7,95/10,95/13,95/14,95/17,95/20,95/23,95/26,95/29,95/32,95/35,95/38,95/41,95/45,95/48,95/51,95/52,95/55,95/58,95/61,95/64,95/67,95/70,95/73,95/76,95/79,95/82,95/85,95/88,95/91,95/94,95/99,96/3,96/6,96/9,96/12,96/13,96/16,96/19,96/22,96/25,96/28,96/31,96/34,96/37,96/40,96/43,96/46,96/51,96/54,96/57,96/60,96/63,96/66,96/69,96/72,96/75,96/78,96/81,96/84,96/87,96/90,96/93,96/97,97/6,97/9,97/12,97/15,97/18,97/21,97/24,97/27,97/30,97/33,97/36,97/39,97/59,97/62,97/65,97/68,97/71,97/74,97/77,97/80,97/83,97/86,97/89,97/92,98/1,98/7,98/23,98/26,98/29,98/32,98/35,98/38,98/41,98/44,98/47,98/52,98/55,98/58,98/61,98/64,98/67,98/70,98/73,98/76,98/79,98/82,98/85,98/88,98/91,98/98,98/99,99/2,99/8,99/25,99/28,99/31,99/34,99/37,99/40,99/43,99/46,99/50,99/53,99/57,99/60,99/63,99/66,99/69,99/72,99/75,99/78,99/81,99/84,99/87,99/90,99/93,99/96,99/99}
\filldraw[color = red] (\x, \y) rectangle (\x+1, \y+1);

%\draw[step=1cm,gray,ultra thin] (0, 0) grid (100,100); 

\draw (0.5 ,-4) node {$0$};
\draw (50.5, -4) node {$50$};
\draw (100.5, -4) node {$100$};

\draw (-5, 0.5) node {$0$};
\draw (-5, 50.5) node {$50$};
\draw (-5, 100.5) node {$100$};

\draw (42, -18) node {Heap 1};
\draw (-23, 40) node {Heap 2};

\end{tikzpicture}
\end{center}
\caption{Initial P-positions of the game given by 
 $\mathcal{M} = \{(0,-2)$, $(-2,0)$, $(2,-3)$, $(-3,2)$, $(-5,4)$, $(-5,-2)$, $(-4,-3)$, $(-1,-4)\}$. Do they eventually settle into a pattern that can be fully understood, or is this a world of ever increasing complexity, where surprises will await us regardless of how far we take our computations?} \label{F:secondExample}
\end{figure}

For many games, the P-positions display some initial irregularities, after which they settle into a simple pattern. However, in some cases, the P-positions show no sign of regularity even after quite extensive computations. Figure~\ref{F:secondExample} shows a typical example with large periodic regions interrupted by rather chaotic behavior along the borders of ``conflicting'' regions. 
  
We claim that despite the apparent simplicity of the rules of these games, it is impossible in general to fully understand the behavior of the set $P(\mathcal{M})$ of P-positions from the list $\mathcal{M}$ of legal moves.
It seems that such a claim requires a formal definition of ``full understanding''. To support our claim, we argue that full understanding, whatever it is, would at least imply the ability to recognize when two games are \emph{P-equivalent}, by which we mean that they have the same P-positions. We will prove that this is algorithmically undecidable.

\begin{Thm} \label{T:main}
There is no algorithm that, given as input the number $d$ of heaps and two finite sets $\mathcal{M}$ and $\mathcal{M}'$ of integer vectors specifying the rules of two $d$-heap games, decides whether or not $P(\mathcal{M}) = P(\mathcal{M}')$. \end{Thm}

The rest of the paper is mainly devoted to the proof of Theorem~\ref{T:main}. In Section~\ref{S:CA} we describe a certain class of cellular automata for which several decision problems are known to be algorithmically unsolvable. In Section~\ref{S:modular} we describe a class of (non-invariant) games called \emph{modular games} that can emulate the cellular automata of Section~\ref{S:CA}. Finally in Section~\ref{S:gadget} we show how invariant games can emulate modular games and how this establishes Theorem~\ref{T:main}.  

\section{Cellular automata} \label{S:CA}
Cellular automata (CAs) give rise to 2-dimensional patterns similar to that of Figure~\ref{F:firstExample}. For CAs it is known that some basic questions are algorithmically undecidable. Here we consider a restricted class of CAs. They have two states (0 and 1), and the state of cell $i$ at time $t$ is denoted by $x_{t,i}$. The starting configuration is $...000111...$, or more precisely, $$x_{0,i} = \begin{cases} 1, \quad \text {if $i\geq 0$},\\ 0, \quad \text{if $i<0$}. \end{cases} $$
The update rule is given by a number $n$ and a Boolean function $f$ taking $n$ bits of input.  The states are updated according to $$x_{t+1,i} = f(x_{t,i-n+1}, x_{t,i-n+2},\dots,x_{t,i}).$$ In other words $x_{t+1,i}$ depends on $x_{t,i}$ and the $n-1$ cells immediately to the left of $x_{t,i}$. Moreover, for technical reasons, we require that $f(0,\dots,0) = 0$, which implies that $x_{t,i} = 0$ whenever $i<0$.

We denote the cellular automaton corresponding to $f$ by $C\!A(f)$.

\subsection*{Example}
We demonstrate by taking $n=2$ and letting the Boolean function be $$f(x,y) = x\oplus y,$$ in other words, $f(x,y)$ is equal to 0 if $x=y$ and 1 if $x\neq y$.
CAs are usually illustrated by drawing the tape from left to right, and for some reason it has become customary to let time flow downwards. Here we break this convention and draw time upwards. The 1's are represented by red squares. There is no need to include the positions $x_{t,i}$ for negative $i$, and therefore the leftmost column represents $x_{t,0}$.

\begin{figure} [h]
\begin{center}
\begin{tikzpicture} [scale = 0.2]

\foreach \x/\y in {0/0, 1/0, 2/0, 3/0, 4/0, 5/0, 6/0, 7/0, 8/0, 9/0, 10/0, 11/0, 12/0, 13/0, 14/0, 15/0, 16/0, 17/0, 18/0, 19/0, 20/0, 21/0, 22/0, 23/0, 24/0, 0/1,0/2,0/3,0/4,0/5,0/6,0/7,0/8,0/9,0/10,0/11,0/12,0/13,0/14,0/15,0/16,0/17,0/18,0/19,0/20,0/21,0/22,0/23,0/24,
1/2,
2/3,
0/4, 1/4, 2/4, 3/4,
4/5,
1/6, 4/6, 5/6,
2/7, 4/7, 6/7,
1/8, 2/8, 3/8, 4/8, 5/8, 6/8, 7/8,
8/9,
1/10,
2/11,
0/12, 1/12, 2/12, 3/12,
4/13,
1/14, 4/14, 5/14,
2/15, 4/15, 6/15,
1/16, 2/16, 3/16, 4/16, 5/16, 6/16, 7/16,
8/10, 9/10,
8/11, 10/11,
8/12, 9/12, 10/12, 11/12,
8/13, 12/13,
8/14, 9/14, 12/14, 13/14,
8/15, 10/15, 12/15, 14/15,
8/16, 9/16, 10/16, 11/16, 12/16, 13/16, 14/16, 15/16,
16/17,
16/18, 17/18,
16/19, 18/19,
16/20, 17/20, 18/20, 19/20,
16/21, 20/21,
16/22, 17/22, 20/22, 21/22,
16/23, 18/23, 20/23, 22/23,
16/24, 17/24, 18/24, 19/24, 20/24, 21/24, 22/24, 23/24,
1/18,
2/19,
1/20, 2/20, 3/20,
4/21,
1/22, 4/22, 5/22,
2/23, 4/23, 6/23,
1/24, 2/24, 3/24, 4/24, 5/24, 6/24, 7/24
}
 \filldraw[color = red] (\x, \y) rectangle (\x+1, \y+1);

\draw[step=1cm,gray,very thin] (0, 0) grid (25,25); 

\draw (0.5 ,-1.2) node {$0$};
\draw (10.5, -1.2) node {$10$};
\draw (20.5, -1.2) node {$20$};

\draw (-1, 0.5) node {$0$};
\draw (-1.5, 10.5) node {$10$};
\draw (-1.5, 20.5) node {$20$};

\draw (12, -4) node {Tape};
\draw (-7, 13) node {Time};

\end{tikzpicture}
\end{center}

\caption{The cellular automaton given by $f(x,y) = x\oplus y$.}
\label{F:CAExample}
\end{figure}
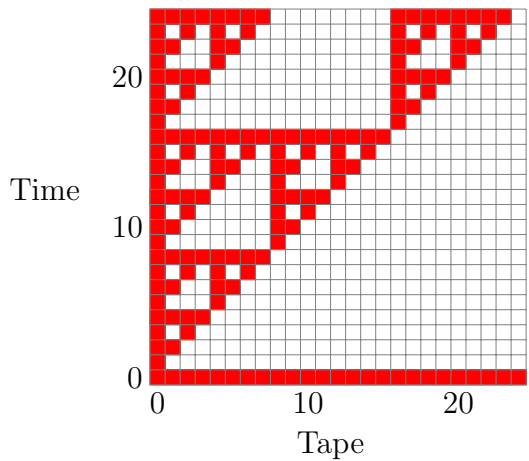

Here the pattern is Pascal's triangle modulo 2, and although this pattern is non-periodic, it is still well understood. However, there are Boolean functions that give rise to more difficult behavior. The following result is well-known.

\begin{Lemma} \label{L:101}
There is no algorithm that takes input an $n$-ary Boolean function $f$, and answers whether or not the string $101$ ever occurs in $C\!A(f)$.
\end{Lemma}

The reason that the lemma is true is essentially that a CA can emulate a generic Turing machine without the substring $101$ occurring in the computation. The CA can then be ``programmed'' to print the string $101$ if the emulated Turing machine halts, making the occurrence of $101$ equivalent to the halting problem \cite{T36}.

The argument works also if the string $101$ is replaced by $00$ or $11$, but $101$ is the simplest string that will do for our subsequent purposes. 

As the similarities between Figures~\ref{F:firstExample} and \ref{F:CAExample} indicate, we will construct a game emulating a generic cellular automaton. %(see also \cite{L2} for simular results)

\section{Modular games}\label{S:modular}
We now describe a class of (non-invariant) games that we call \emph{modular} games. We show that an arbitrary CA can be emulated by a modular game, and that the 101-occurrence problem can be reduced to P-equivalence of modular games. 

A modular game $\mathcal{G}$ has only two heaps (which we call the tape- and the time-heap because they correspond to the tape- and time-axes in Figure~\ref{F:CAExample}). 
For some positive integer $k$, there are finite sets $\mathcal{M}_0,\dots, \mathcal{M}_{k-1}$ of integer vectors that specify the available move options in $\mathcal{G}$. In a given position $(a_1, a_2)$, where $a_1$ is the number of matches in the tape-heap and $a_2$ is the number of matches in the time-heap, the set of available moves is given by $\mathcal{M}_i$, where $0\leq i < k$ and 
\begin{align}\label{ia2}
i\equiv a_2 \pmod k. 
\end{align}
Moreover we require that the number of matches in the time-heap decreases in every move and that the number of matches in the tape-heap does not increase. That is if $(m_1,m_2)\in \mathcal{M}_i$ then $m_1\le 0$ and $m_2<0$ and there is a move option from $(a_1,a_2)$ to $(a_1+m_1, a_2+m_2)$ provided $a_1+m_1\ge 0, a_2+m_2\ge 0$ and (\ref{ia2}) hold. We wish to prove the following lemma.

\begin{Lemma} \label{L:101->modular}
For each Boolean function $f$ we can effectively construct two modular games $\mathcal{G}$ and $\mathcal{H}$ that are P-equivalent if and only if the word $101$ never occurs in $C\!A(f)$.
\end{Lemma}

The first step is to construct a modular game that emulates $C\!A(f)$ for an arbitrary Boolean function $f$.

\subsection{The modular games and computation}
The number of matches in the tape-heap will correspond in the obvious way to a position on the tape. On the time axis we insert some space to allow for the rules of the modular game to ``compute'' the Boolean function $f$. 

For some $k$ which will have to depend on $f$, the positions with $kt$ matches in the time-heap will correspond to the state of $C\!A(f)$ at time $t$. Hence the positions where the number of matches in the time-heap is divisible by $k$ will encode the evolution of $C\!A(f)$.

Let square brackets $[\cdot]$ denote $1-\max(\cdot)$, so that $$\left[x y z\right] = 1 - \max(x, y, z),$$  etc, and by convention $[\, ] = 1$. It is well-known that every Boolean function can be expressed in terms of nested brackets. This is because the set consisting of \& and $\sim$ (negation) is a complete set of connectives in propositional logic. For instance, as can easily be checked, the function $f(x,y)=x\oplus y$ can be expressed as 
$$x\oplus y = [[xy][[x][y]]].$$
If we let P-positions correspond to the value 1 and N-positions (the non-P positions) to the value 0, then the value of a position can be recursively computed by the bracket-function. If a position has moves to positions of values $x_1,\dots,x_k$, then its value is $[x_1\cdots x_k]$.

\begin{figure} [h]
\begin{center}
\begin{tikzpicture} [scale = 0.6]

\foreach \x/\y in {4/0, 5/0, 5/1, 5/2, 5/3, 5/4, 5/5}
 \draw[thick] (\x, \y) rectangle (\x+1, \y+1);

\draw (0 , 0.5) node {$x$};
\draw (1, 0.5) node {$y$};
\draw (1, 1.5) node {$[x]$};
\draw (1, 2.5) node {$[y]$};
\draw (1, 3.5) node {$[xy]$};
\draw (1, 4.5) node {$[[x][y]]$};
\draw (1, 5.5) node {$[[xy][[x][y]]]$};

\draw[ultra thick, ->] (5, 1.4) ..controls (4.7, 1.4) and (4.6, 1.3) .. (4.6, 1);
\draw[ultra thick, ->] (5, 3.5) ..controls (4.4, 3.3) .. (4.3, 1);

\draw[ultra thick, ->] (6, 2.4) ..controls (7, 2.4) and (7, 0.7) .. (6, 0.7);
\draw[ultra thick, ->] (6, 3.5) ..controls (8, 3.5) and (8, 0.3) .. (6, 0.3);
\draw[ultra thick, ->] (6, 4.5) ..controls (7, 4.5) and (7, 2.7) .. (6, 2.7);

\draw[ultra thick, ->] (5, 4.3) ..controls (4, 4.3) and (4, 1.7) .. (5, 1.7);
\draw[ultra thick, ->] (5, 5.6) ..controls (4, 5.5) and (4, 3.8) .. (5, 3.7);
\draw[ultra thick, ->] (5, 5.4) ..controls (4.5, 5.4) and (4.5, 4.6) .. (5, 4.6);

\end{tikzpicture}
\end{center}

\caption{A modular game computing $f(x,y) = x\oplus y$ in five steps. The arrows indicate move options. The value of each cell is the bracket of the values of all its options.}
\label{F:xor}
\end{figure}
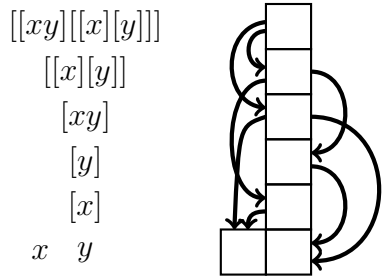

Therefore we can construct a game that computes the function $f$ by letting intermediate positions have values $[x]$, $[y]$, $[xy]$ and $[[x][y]]$. Figure~\ref{F:xor} illustrates how this is done for $k=5$.

The construction in Figure~\ref{F:xor} corresponds to 
\begin{eqnarray}
\notag \mathcal{M}_0 &=& \{(0,-1), (0,-2)\} \\
\notag \mathcal{M}_1 &=& \{(-1,-1)\} \\
\notag \mathcal{M}_2 &=& \{(0,-2)\} \\
\notag \mathcal{M}_3 &=& \{(0,-3),(-1,-3)\} \\
\notag \mathcal{M}_4 &=& \{(0,-2), (0,-3)\}
\end{eqnarray}

\begin{figure} [h]
\begin{center}
\begin{tikzpicture} [scale = 0.15]

\foreach \x/\y in {
0/0,0/1,0/5,0/6,0/10,0/11,0/15,0/16,0/20,0/21,0/25,0/26,0/30,0/31,0/35,0/36,0/40,0/41,0/45,0/46,1/0,1/4,1/7,1/10,1/14,1/17,1/20,1/24,1/27,1/30,1/34,1/37,1/40,1/44,1/47,2/0,2/4,2/6,2/7,2/8,2/12,2/15,2/16,2/20,2/24,2/26,2/27,2/28,2/32,2/35,2/36,2/40,2/44,2/46,2/47,2/48,3/0,3/4,3/6,3/7,3/8,3/11,3/12,3/13,3/17,3/20,3/24,3/26,3/27,3/28,3/31,3/32,3/33,3/37,3/40,3/44,3/46,3/47,3/48,4/0,4/4,4/6,4/7,4/8,4/11,4/12,4/13,4/16,4/17,4/18,4/22,4/25,4/26,4/30,4/31,4/35,4/36,4/40,4/44,4/46,4/47,4/48,5/0,5/4,5/6,5/7,5/8,5/11,5/12,5/13,5/16,5/17,5/18,5/21,5/22,5/23,5/27,5/30,5/34,5/37,5/40,5/44,5/46,5/47,5/48,6/0,6/4,6/6,6/7,6/8,6/11,6/12,6/13,6/16,6/17,6/18,6/21,6/22,6/23,6/26,6/27,6/28,6/32,6/35,6/36,6/40,6/44,6/46,6/47,6/48,7/0,7/4,7/6,7/7,7/8,7/11,7/12,7/13,7/16,7/17,7/18,7/21,7/22,7/23,7/26,7/27,7/28,7/31,7/32,7/33,7/37,7/40,7/44,7/46,7/47,7/48,8/0,8/4,8/6,8/7,8/8,8/11,8/12,8/13,8/16,8/17,8/18,8/21,8/22,8/23,8/26,8/27,8/28,8/31,8/32,8/33,8/36,8/37,8/38,8/42,8/45,8/46,9/0,9/4,9/6,9/7,9/8,9/11,9/12,9/13,9/16,9/17,9/18,9/21,9/22,9/23,9/26,9/27,9/28,9/31,9/32,9/33,9/36,9/37,9/38,9/41,9/42,9/43,9/47,10/0,10/4,10/6,10/7,10/8,10/11,10/12,10/13,10/16,10/17,10/18,10/21,10/22,10/23,10/26,10/27,10/28,10/31,10/32,10/33,10/36,10/37,10/38,10/41,10/42,10/43,10/46,10/47,10/48,11/0,11/4,11/6,11/7,11/8,11/11,11/12,11/13,11/16,11/17,11/18,11/21,11/22,11/23,11/26,11/27,11/28,11/31,11/32,11/33,11/36,11/37,11/38,11/41,11/42,11/43,11/46,11/47,11/48,12/0,12/4,12/6,12/7,12/8,12/11,12/12,12/13,12/16,12/17,12/18,12/21,12/22,12/23,12/26,12/27,12/28,12/31,12/32,12/33,12/36,12/37,12/38,12/41,12/42,12/43,12/46,12/47,12/48,13/0,13/4,13/6,13/7,13/8,13/11,13/12,13/13,13/16,13/17,13/18,13/21,13/22,13/23,13/26,13/27,13/28,13/31,13/32,13/33,13/36,13/37,13/38,13/41,13/42,13/43,13/46,13/47,13/48,14/0,14/4,14/6,14/7,14/8,14/11,14/12,14/13,14/16,14/17,14/18,14/21,14/22,14/23,14/26,14/27,14/28,14/31,14/32,14/33,14/36,14/37,14/38,14/41,14/42,14/43,14/46,14/47,14/48,15/0,15/4,15/6,15/7,15/8,15/11,15/12,15/13,15/16,15/17,15/18,15/21,15/22,15/23,15/26,15/27,15/28,15/31,15/32,15/33,15/36,15/37,15/38,15/41,15/42,15/43,15/46,15/47,15/48,16/0,16/4,16/6,16/7,16/8,16/11,16/12,16/13,16/16,16/17,16/18,16/21,16/22,16/23,16/26,16/27,16/28,16/31,16/32,16/33,16/36,16/37,16/38,16/41,16/42,16/43,16/46,16/47,16/48,17/0,17/4,17/6,17/7,17/8,17/11,17/12,17/13,17/16,17/17,17/18,17/21,17/22,17/23,17/26,17/27,17/28,17/31,17/32,17/33,17/36,17/37,17/38,17/41,17/42,17/43,17/46,17/47,17/48,18/0,18/4,18/6,18/7,18/8,18/11,18/12,18/13,18/16,18/17,18/18,18/21,18/22,18/23,18/26,18/27,18/28,18/31,18/32,18/33,18/36,18/37,18/38,18/41,18/42,18/43,18/46,18/47,18/48,19/0,19/4,19/6,19/7,19/8,19/11,19/12,19/13,19/16,19/17,19/18,19/21,19/22,19/23,19/26,19/27,19/28,19/31,19/32,19/33,19/36,19/37,19/38,19/41,19/42,19/43,19/46,19/47,19/48,20/0,20/4,20/6,20/7,20/8,20/11,20/12,20/13,20/16,20/17,20/18,20/21,20/22,20/23,20/26,20/27,20/28,20/31,20/32,20/33,20/36,20/37,20/38,20/41,20/42,20/43,20/46,20/47,20/48,21/0,21/4,21/6,21/7,21/8,21/11,21/12,21/13,21/16,21/17,21/18,21/21,21/22,21/23,21/26,21/27,21/28,21/31,21/32,21/33,21/36,21/37,21/38,21/41,21/42,21/43,21/46,21/47,21/48,22/0,22/4,22/6,22/7,22/8,22/11,22/12,22/13,22/16,22/17,22/18,22/21,22/22,22/23,22/26,22/27,22/28,22/31,22/32,22/33,22/36,22/37,22/38,22/41,22/42,22/43,22/46,22/47,22/48,23/0,23/4,23/6,23/7,23/8,23/11,23/12,23/13,23/16,23/17,23/18,23/21,23/22,23/23,23/26,23/27,23/28,23/31,23/32,23/33,23/36,23/37,23/38,23/41,23/42,23/43,23/46,23/47,23/48,24/0,24/4,24/6,24/7,24/8,24/11,24/12,24/13,24/16,24/17,24/18,24/21,24/22,24/23,24/26,24/27,24/28,24/31,24/32,24/33,24/36,24/37,24/38,24/41,24/42,24/43,24/46,24/47,24/48,25/0,25/4,25/6,25/7,25/8,25/11,25/12,25/13,25/16,25/17,25/18,25/21,25/22,25/23,25/26,25/27,25/28,25/31,25/32,25/33,25/36,25/37,25/38,25/41,25/42,25/43,25/46,25/47,25/48,26/0,26/4,26/6,26/7,26/8,26/11,26/12,26/13,26/16,26/17,26/18,26/21,26/22,26/23,26/26,26/27,26/28,26/31,26/32,26/33,26/36,26/37,26/38,26/41,26/42,26/43,26/46,26/47,26/48,27/0,27/4,27/6,27/7,27/8,27/11,27/12,27/13,27/16,27/17,27/18,27/21,27/22,27/23,27/26,27/27,27/28,27/31,27/32,27/33,27/36,27/37,27/38,27/41,27/42,27/43,27/46,27/47,27/48,28/0,28/4,28/6,28/7,28/8,28/11,28/12,28/13,28/16,28/17,28/18,28/21,28/22,28/23,28/26,28/27,28/28,28/31,28/32,28/33,28/36,28/37,28/38,28/41,28/42,28/43,28/46,28/47,28/48,29/0,29/4,29/6,29/7,29/8,29/11,29/12,29/13,29/16,29/17,29/18,29/21,29/22,29/23,29/26,29/27,29/28,29/31,29/32,29/33,29/36,29/37,29/38,29/41,29/42,29/43,29/46,29/47,29/48,30/0,30/4,30/6,30/7,30/8,30/11,30/12,30/13,30/16,30/17,30/18,30/21,30/22,30/23,30/26,30/27,30/28,30/31,30/32,30/33,30/36,30/37,30/38,30/41,30/42,30/43,30/46,30/47,30/48,31/0,31/4,31/6,31/7,31/8,31/11,31/12,31/13,31/16,31/17,31/18,31/21,31/22,31/23,31/26,31/27,31/28,31/31,31/32,31/33,31/36,31/37,31/38,31/41,31/42,31/43,31/46,31/47,31/48,32/0,32/4,32/6,32/7,32/8,32/11,32/12,32/13,32/16,32/17,32/18,32/21,32/22,32/23,32/26,32/27,32/28,32/31,32/32,32/33,32/36,32/37,32/38,32/41,32/42,32/43,32/46,32/47,32/48,33/0,33/4,33/6,33/7,33/8,33/11,33/12,33/13,33/16,33/17,33/18,33/21,33/22,33/23,33/26,33/27,33/28,33/31,33/32,33/33,33/36,33/37,33/38,33/41,33/42,33/43,33/46,33/47,33/48,34/0,34/4,34/6,34/7,34/8,34/11,34/12,34/13,34/16,34/17,34/18,34/21,34/22,34/23,34/26,34/27,34/28,34/31,34/32,34/33,34/36,34/37,34/38,34/41,34/42,34/43,34/46,34/47,34/48,35/0,35/4,35/6,35/7,35/8,35/11,35/12,35/13,35/16,35/17,35/18,35/21,35/22,35/23,35/26,35/27,35/28,35/31,35/32,35/33,35/36,35/37,35/38,35/41,35/42,35/43,35/46,35/47,35/48,36/0,36/4,36/6,36/7,36/8,36/11,36/12,36/13,36/16,36/17,36/18,36/21,36/22,36/23,36/26,36/27,36/28,36/31,36/32,36/33,36/36,36/37,36/38,36/41,36/42,36/43,36/46,36/47,36/48,37/0,37/4,37/6,37/7,37/8,37/11,37/12,37/13,37/16,37/17,37/18,37/21,37/22,37/23,37/26,37/27,37/28,37/31,37/32,37/33,37/36,37/37,37/38,37/41,37/42,37/43,37/46,37/47,37/48,38/0,38/4,38/6,38/7,38/8,38/11,38/12,38/13,38/16,38/17,38/18,38/21,38/22,38/23,38/26,38/27,38/28,38/31,38/32,38/33,38/36,38/37,38/38,38/41,38/42,38/43,38/46,38/47,38/48,39/0,39/4,39/6,39/7,39/8,39/11,39/12,39/13,39/16,39/17,39/18,39/21,39/22,39/23,39/26,39/27,39/28,39/31,39/32,39/33,39/36,39/37,39/38,39/41,39/42,39/43,39/46,39/47,39/48,40/0,40/4,40/6,40/7,40/8,40/11,40/12,40/13,40/16,40/17,40/18,40/21,40/22,40/23,40/26,40/27,40/28,40/31,40/32,40/33,40/36,40/37,40/38,40/41,40/42,40/43,40/46,40/47,40/48,41/0,41/4,41/6,41/7,41/8,41/11,41/12,41/13,41/16,41/17,41/18,41/21,41/22,41/23,41/26,41/27,41/28,41/31,41/32,41/33,41/36,41/37,41/38,41/41,41/42,41/43,41/46,41/47,41/48,42/0,42/4,42/6,42/7,42/8,42/11,42/12,42/13,42/16,42/17,42/18,42/21,42/22,42/23,42/26,42/27,42/28,42/31,42/32,42/33,42/36,42/37,42/38,42/41,42/42,42/43,42/46,42/47,42/48,43/0,43/4,43/6,43/7,43/8,43/11,43/12,43/13,43/16,43/17,43/18,43/21,43/22,43/23,43/26,43/27,43/28,43/31,43/32,43/33,43/36,43/37,43/38,43/41,43/42,43/43,43/46,43/47,43/48,44/0,44/4,44/6,44/7,44/8,44/11,44/12,44/13,44/16,44/17,44/18,44/21,44/22,44/23,44/26,44/27,44/28,44/31,44/32,44/33,44/36,44/37,44/38,44/41,44/42,44/43,44/46,44/47,44/48,45/0,45/4,45/6,45/7,45/8,45/11,45/12,45/13,45/16,45/17,45/18,45/21,45/22,45/23,45/26,45/27,45/28,45/31,45/32,45/33,45/36,45/37,45/38,45/41,45/42,45/43,45/46,45/47,45/48,46/0,46/4,46/6,46/7,46/8,46/11,46/12,46/13,46/16,46/17,46/18,46/21,46/22,46/23,46/26,46/27,46/28,46/31,46/32,46/33,46/36,46/37,46/38,46/41,46/42,46/43,46/46,46/47,46/48,47/0,47/4,47/6,47/7,47/8,47/11,47/12,47/13,47/16,47/17,47/18,47/21,47/22,47/23,47/26,47/27,47/28,47/31,47/32,47/33,47/36,47/37,47/38,47/41,47/42,47/43,47/46,47/47,47/48,48/0,48/4,48/6,48/7,48/8,48/11,48/12,48/13,48/16,48/17,48/18,48/21,48/22,48/23,48/26,48/27,48/28,48/31,48/32,48/33,48/36,48/37,48/38,48/41,48/42,48/43,48/46,48/47,48/48,49/0,49/4,49/6,49/7,49/8,49/11,49/12,49/13,49/16,49/17,49/18,49/21,49/22,49/23,49/26,49/27,49/28,49/31,49/32,49/33,49/36,49/37,49/38,49/41,49/42,49/43,49/46,49/47,49/48
}
 \filldraw[color = gray, opacity = 0.6] (\x, \y) rectangle (\x+1, \y+1);

 \foreach \x/\y in {
0/0,0/5,0/10,0/15,0/20,0/25,0/30,0/35,0/40,0/45,1/0,1/10,1/20,1/30,1/40,2/0,2/15,2/20,2/35,2/40,3/0,3/20,3/40,4/0,4/25,4/30,4/35,4/40,5/0,5/30,5/40,6/0,6/35,6/40,7/0,7/40,8/0,8/45,9/0,10/0,11/0,12/0,13/0,14/0,15/0,16/0,17/0,18/0,19/0,20/0,21/0,22/0,23/0,24/0,25/0,26/0,27/0,28/0,29/0,30/0,31/0,32/0,33/0,34/0,35/0,36/0,37/0,38/0,39/0,40/0,41/0,42/0,43/0,44/0,45/0,46/0,47/0,48/0,49/0
}
\filldraw[color = red] (\x, \y) rectangle (\x+1, \y+1);

\draw[step=1cm,gray,very thin] (0, 0) grid (50,50); 

\draw (25, -4) node {Tape};
\draw (-7, 25) node {Time};

\end{tikzpicture}
\end{center}

\caption{A modular game emulating $f(x,y) = x\oplus y$. Here the P-positions with fewer than 50 matches in each heap are represented by filled squares. Rows corresponding to $a_2\equiv 0$ (mod 5) are highlighted by drawing the P-positions in red.}
\label{F:modularExample}
\end{figure}
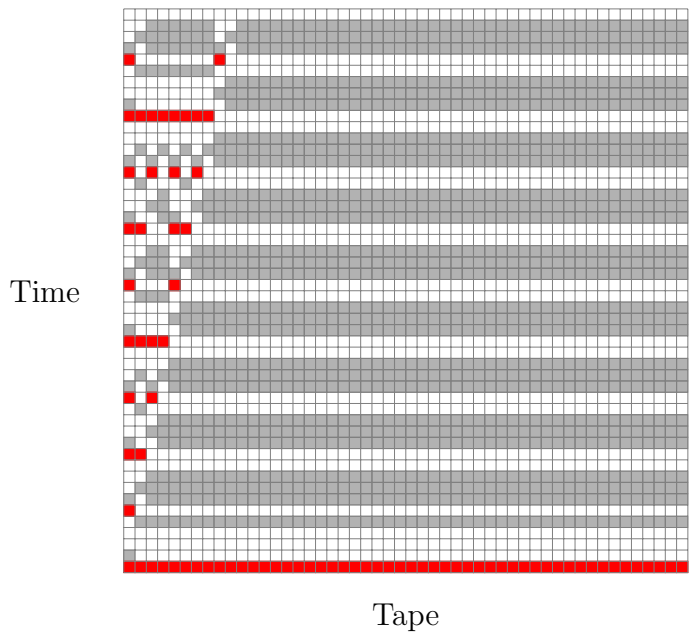

Figure~\ref{F:modularExample} shows how this modular game emulates the cellular automaton given by $f(x,y) = x \oplus y$. Every fifth row corresponds to a row of Figure~\ref{F:CAExample}.

\subsection{The check for $101$}
The second step concerns the check for the word $101$. 
Given a modular game $\mathcal{G}$, corresponding to a Boolean function $f$ as illustrated in Figure~\ref{F:xor}, we can construct a modified game where, inside the computation of $f$, we have built in a check of whether or not the input to $f$ ends with the substring $101$. 

The check takes place before the actual computation of $f$. First we introduce boxes $b_1$ and $b_2$ that invert the last position and the third position from the end of the input (which is where we want to check if there are 1's in order to check if the input ends by 101). Then the box $b_3$ that actually checks for $101$ is connected to the penultimate position in the input, and to $b_1$ and $b_2$. We denote this game by $\mathcal{G}'$.

We also construct a ``dummy'' game $\mathcal{G}''$ that looks like $\mathcal{G}'$ but doesn't check for $101$. In $\mathcal{G}''$, the box $b_3$ is instead connected only to $b_2$ and to the third box from the end of the input. Therefore the box $b_3$ will always be an N-position (a zero) in $\mathcal{G}''$, while in $\mathcal{G}'$ it will be a P-position whenever the input string ends by 101. Figures~\ref{F:G'} and \ref{F:G''} depict the construction of the games $\mathcal{G}'$ and $\mathcal{G}''$. 

Hence, the modular games $\mathcal{G}'$ and $\mathcal{G}''$ are P-equivalent if and only if the string $101$ never occurs in $C\!A(f)$. This proves Lemma~\ref{L:101->modular}.

\begin{figure} [h]
\begin{center}
\begin{tikzpicture} [scale = 0.6]

\foreach \x/\y in {3/0, 4/0, 5/0, 5/1, 5/2, 5/3, 5/5}
 \draw[thick] (\x, \y) rectangle (\x+1, \y+1);

\draw (3.5 , 0.5) node {$1$};
\draw (4.5, 0.5) node {$0$};
\draw (5.5, 0.5) node {$1$};

\draw (5.5, 1.5) node {$0$};
\draw (5.5, 2.5) node {$0$};
\draw (5.5, 3.5) node {$1$};
\draw (0, 5.5) node {$f$};
\draw (-2 , 0.5) node {$\ldots$};
\draw (-1 , 0.5) node {$x$};
\draw (-.2, 0.5) node {$y$};
\draw (0.6, 0.5) node {$z$};
\draw (2.3, 0.5) node {$\ldots$};
\draw (0, 1.5) node {$b_1=[z]$};
\draw (0, 2.5) node {$b_2=[x]$};
\draw (0, 3.5) node {$b_3$};
\draw (0, 4.3) node {$.$};
\draw (0, 4.5) node {$.$};
\draw (0, 4.7) node {$.$};
\draw (5.5, 4.3) node {$.$};
\draw (5.5, 4.5) node {$.$};
\draw (5.5, 4.7) node {$.$};

\draw[ultra thick, ->] (5, 3.5) ..controls (4.4, 3.5) and (4.5, 1.3) .. (4.5, 1);
\draw[ultra thick, ->] (5, 2.5) ..controls (3.5, 2.5) and (3.5, 1.4) .. (3.5, 1);
\draw[ultra thick, ->] (6, 1.3) ..controls (6.8, 1.3) and (6.8, 0.5) .. (6, 0.5);
\draw[ultra thick, ->] (6, 3.7) ..controls (7.3, 3.7) and (7.3, 1.7) .. (6, 1.7);
\draw[ultra thick, ->] (6, 3.3) ..controls (6.8, 3.3) and (6.8, 2.5) .. (6, 2.5);

\end{tikzpicture}
%\end{center}

\caption{A modular game $\mathcal{G}'$ with input $xyz$ computing a function $f$ in a few steps. (We have omitted the moves that actually compute $f$ and any remaining variables.) The first two boxes $b_1$ and $b_2$ invert $x$ and $z$. The third box $b_3$ is a $1$ (=P-position) if and only if $xyz = 101$.}
\label{F:G'}
\vspace{0.5 cm}
\begin{tikzpicture} [scale = 0.6]

\foreach \x/\y in {3/0, 4/0, 5/0, 5/1, 5/2, 5/3, 5/5}
 \draw[thick] (\x, \y) rectangle (\x+1, \y+1);

\draw (3.5 , 0.5) node {$1$};
\draw (4.5, 0.5) node {$0$};
\draw (5.5, 0.5) node {$1$};

\draw (5.5, 1.5) node {$0$};
\draw (5.5, 2.5) node {$0$};
\draw (5.5, 3.5) node {$0$};
\draw (-2, 0.5) node {$\ldots$};
\draw (-1 , 0.5) node {$x$};
\draw (-.2, 0.5) node {$y$};
\draw (.6, 0.5) node {$z$};
\draw (2.3, 0.5) node {$\ldots$};
\draw (0, 1.5) node {$b_1=[z]$};
\draw (0, 2.5) node {$b_2=[x]$};
\draw (0, 3.5) node {$b_3=0$};
\draw (0, 5.5) node {$f$};

\draw (0, 4.3) node {$.$};
\draw (0, 4.5) node {$.$};
\draw (0, 4.7) node {$.$};
\draw (5.5, 4.3) node {$.$};
\draw (5.5, 4.5) node {$.$};
\draw (5.5, 4.7) node {$.$};

\draw[ultra thick, ->] (5, 2.5) ..controls (3.5, 2.5) and (3.7, 1.4) .. (3.7, 1);
\draw[ultra thick, ->] (6, 3.5) ..controls (6.8, 3.5) and (6.8, 2.5) .. (6, 2.5);
\draw[ultra thick, ->] (6, 1.5) ..controls (6.8, 1.5) and (6.8, 0.5) .. (6, 0.5);
\draw[ultra thick, ->] (5, 3.5) ..controls (3, 3.5) and (3.3, 1.3) .. (3.3, 1);
%%%%%%%%%%%%%%%%%%%

\end{tikzpicture}
\end{center}

\caption{A modular game $\mathcal{G}''$ whose output function $f$ is identical to that of $\mathcal{G}'$. The only difference is that $\mathcal{G}''$ does not check for the pattern $101$. The box $b_3$ contains a $0$ independently of the input $xyz$ to $f$.}\label{F:G''}

\end{figure}

\section{Emulating a modular game by an invariant game} \label{S:gadget}
The following lemma, which we prove in this section, provides the remaining link to Theorem \ref{T:main}. 
 
\begin{Lemma}\label{L:eqPeq}
For two modular games $\mathcal{G}$ and $\mathcal{H}$, we can effectively construct invariant games $\mathcal{M}$ and $\mathcal{M}'$ that are P-equivalent if and only if $\mathcal{G}$ and $\mathcal{H}$ are P-equivalent.
\end{Lemma}

We introduce a gadget that allows us to emulate a modular game by an invariant game. The resulting invariant game has a time-heap and a tape-heap, and $k$ more heaps that we call the \emph{gadget}, keeping track of the size of the time-heap modulo $k$. The gadget consists of $k$ heaps where the positions that we are primarily interested in are those where one of the heaps contains a match and the remaining $k-1$ heaps are empty (we return to what happens if this is not the case). The moves are designed so that the match in the gadget follows the time-heap modulo $k$. The heaps of the gadget may be numbered $0,\dots , k-1$, and for each of the move sets $\mathcal{M}_i$ of the modular game, we introduce corresponding moves in the invariant game where the $i$th heap of the gadget is emptied, the tape- and time-heaps are affected as given by some move in $\mathcal{M}_i$, and a match is added to the heap of the gadget corresponding to the new modulus of the time-heap.

There is a small technicality to address here. In order for the construction to work, we must require that each move in the modular game changes the congruence class of the number of matches in the time-heap. But this constraint is taken care of by replacing $k$ by a multiple of $k$ which is larger than the number of matches removed from the time-heap in any of the moves (a $k$-modular game can be described as an $lk$-modular game for every positive integer $l$). 

Clearly there is a subset of positions of the invariant game that emulate the modular game, namely those where (i) there is exactly one match in the gadget, and (ii) this match is in the heap corresponding to the number of matches in the time-heap modulo $k$. 

\subsection{Reducing pattern-occurrence to P-equivalence}

In order to conclude that the question of P-equivalence of invariant games is algorithmically undecidable, we wish to reduce the pattern occurrence problem of a generic CA to P-equivalence. We did this for modular games in Section~\ref{S:modular}. It remains to check that the result actually carries over to invariant games through the gadget-trick. 

\subsubsection{If the gadget is out of phase}

In case a single match in the gadget is out of phase with the congruence class of the time-heap, there is a first row $0<i<k$ which corresponds to a finished computation as in Figure~\ref{F:xor}. The gadget treats this row as if it were congruent to 0 modulo $k$ (that is permitting moves as defined by $\mathcal{M}_0$). But, since $i-k<0$, by the convention $[\,] = 1$, $f(x_1,\ldots ,x_n)$ is defined uniquely by its empty input producing an output which is independent of the position of the tape-heap. Hence, the information in row $i$ must be constant. If it is constant 0 (N-positions), then the following pattern will be periodic, since the computation now restarts as if it had started on a tape of only zeros. If on the other hand it is constant 1 (P-positions), then the behavior from row $i$ and onwards will be the same as if the gadget was in phase, since the computation now starts from a row identical to that of the starting configuration of the CA, that is $\cdots 000111\cdots$.

\subsubsection{If the gadget contains more than one match}
What happens in positions where the gadget contains more than one match? 
Say that there is a match in heap $i$ of the gadget and another in heap $j\ne i$. Then a legal move in the time-heap can be obtained from either $\mathcal{M}_i$ or $\mathcal{M}_j$ independent of its number of matches, so that the rules of the invariant game will not encode the evolution of $C\!A(f)$ as prescribed by the modular games in Section \ref{S:modular}. Hence we would like to make positions with more than one match in the gadget trivial. We therefore choose some large number $N$, and allow any move that transfers two matches in the gadget to two other heaps, and removes any number smaller than $N$ from the two main heaps. Since the number of matches in the gadget will never change, this will give a trivial periodic pattern of P-positions in the main heaps. 

\section{Conclusion and questions}
In conclusion, a sub-class of all invariant games emulate the modular games, for which, by  Lemma \ref{L:101} and Lemma \ref{L:101->modular}, it is undecidable whether or not two games are $P$-equivalent. Altogether, this proves Lemma \ref{L:eqPeq} and hence also Theorem \ref{T:main}.
As a consequence of our approach it is also algorithmically undecidable whether a given finite pattern occurs in the set of P-positions of an invariant game. 

How many heaps are required for undecidability? (Strictly speaking we didn't prove that \emph{any} finite number of heaps leads to undecidability). We guess that three heaps suffice, and perhaps even two since it is easy to generate complicated patterns of P-positions as in Figure \ref{F:secondExample} with small finite sets of moves.

Do we need to be able to add matches to heaps in order to achieve undecidability?

If there are no restrictions on moves, so that the total number of matches may increase, is the outcome  (P, N or draw) of a specific position decidable?


\begin{thebibliography}{11}

\bibitem[BCG04]{BCG04} E.\ R.\ Berlekamp, J.\ H.\ Conway, R.\ K.\ Guy, 
\emph{Winning ways}, {\bf 1-2} Academic Press, London (1982). Second edition, {\bf 1-4}. A. K. Peters, Wellesley/MA (2001/03/03/04).

\bibitem[DR10]{DR10} E.\ Duch\^{e}ne and M. Rigo, Invariant Games,
\emph{Theoret. Comput. Sci.}, Vol. 411, 34-36 (2010), pp. 3169--3180 

\bibitem[G66]{G66} S.\ W.\ Golomb, A mathematical investigation of games of 
``take-away''. \emph{J. Combinatorial Theory} {\bf 1} (1966) pp. 443--458.

\bibitem[L12]{L12} U.\ Larsson, The $\star$-operator and invariant subtraction games, \emph{Theoret. Comput. Sci.}, Vol. 422, (2012) pp. 52--58. %http://dx.doi.org/10.1016/j.tcs.2011.11.021.

\bibitem[LHF11]{LHF11} U.\ Larsson, P.\ Hegarty, A.\ S.\ Fraenkel, Invariant 
and dual subtraction games resolving the Duch\^ene-Rigo Conjecture, 
\emph{Theoret. Comp. Sci.} Vol. 412, 8-10 (2011) pp. 729--735.

\bibitem[T36]{T36} A.\ M.\ Turing, On Computable Numbers, with an Application to the Entscheidungsproblem. \emph{Proc. London Math. Soc., series 2}, {\bf 42} (1936-37), pp. 230--265.

\bibitem[W02]{W02} S.\ Wolfram, \emph{A New Kind of Science}, Champaign, IL: Wolfram Media, Inc., (2002).

\end{thebibliography}
\end{document}